\documentclass{aa}  
\usepackage{graphicx,upgreek,txfonts}
\begin{document}

   \title{Long X-ray flares from the central source in RCW\,103}
 \subtitle{\emph{XMM--Newton} and VLT observations in the aftermath of the 2016 outburst}
\titlerunning{Long flares from \mbox{1E\,161348--5055}}
\authorrunning{P. Esposito et al.}

   \author{P.~Esposito,\inst{1} A.~De~Luca,\inst{1,2} R.~Turolla,\inst{3,4}
   F.~Coti~Zelati,\inst{5,6} W.~Hummel,\inst{7} A.~Tiengo,\inst{8,1,2} G.\,L.~Israel,\inst{9} N.~Rea,\inst{5,6} R.\,P.~Mignani,\inst{1,10} \and A.~Borghese\inst{11}
          }

   \institute{INAF--Istituto di Astrofisica Spaziale e Fisica Cosmica di Milano, 
      via A. Corti 12, 20133 Milano, Italy\\
              \email{paolo.esposito@inaf.it}
  \and INFN--Istituto Nazionale di Fisica Nucleare, Sezione di 
  Pavia, via A. Bassi 6, 27100 Pavia, Italy
              \and
              Dipartimento di Fisica e Astronomia `Galileo Galilei', Universit\`a di Padova, via F. Marzolo 8, 35131 Padova, Italy
              \and Mullard Space Science Laboratory, University College London, Holmbury St. Mary, Dorking, Surrey RH5 6NT, UK
              \and Institute of Space Sciences (ICE, CSIC), Campus UAB, Carrer de Can
Magrans s/n, 08193 Barcelona, Spain
              \and Institut d'Estudis Espacials de Catalunya (IEEC), Gran Capit\`a 2--4,
08034 Barcelona, Spain
              \and European Southern Observatory, Karl-Schwarzschild-Stra\ss{}e 2, 85748 Garching bei M\"unchen, Germany
              \and Scuola Universitaria Superiore IUSS Pavia, 
      piazza della Vittoria 15, 27100 Pavia, Italy
              \and INAF--Osservatorio Astronomico di Roma, via Frascati 33, 00078 Monteporzio Catone, Italy
              \and Janusz Gil Institute of Astronomy, University of Zielona G\'ora, ul. Szafrana 2, 65--516, Zielona G\'ora, Poland
              \and Anton Pannekoek Institute for Astronomy, University of Amsterdam, Science Park 904, 1098 XH Amsterdam, The Netherlands
      }

  \date{Received 06 March 2019 / accepted 10 April 2019}

  \abstract{We observed the slowly revolving pulsar \mbox{1E\,161348--5055} (\mbox{1E\,1613}, spin period of 6.67\,h) in the supernova remnant RCW\,103 twice with \emph{XMM--Newton} and once with the Very Large Telescope (VLT). The VLT observation was performed on 2016 June 30, about a week after the detection of a large outburst from \mbox{1E\,1613}. At the position of \mbox{1E\,1613}, we found a near-infrared source with $K_S=20.68\pm0.12$\,mag that was not detected ($K_S>21.2$\,mag) in data collected with the same instruments in 2006, during X-ray quiescence. Its position and behavior are consistent with a counterpart in the literature that was discovered with the \emph{Hubble Space Telescope} in the following weeks in adjacent near-IR bands. 
  The \emph{XMM--Newton} pointings were carried out on 2016 August 19 and on 2018 February 14. While the collected spectra are similar in shape between each other and to what is observed in quiescence (a blackbody with $kT\sim0.5$\,keV plus a second, harder component, either another hotter blackbody with $kT\sim1.2$\,keV or a power law with photon index $\Gamma\sim3$), the two pointings caught \mbox{1E\,1613} at different luminosity throughout its decay pattern: about $4.8\times10^{34}$\,\mbox{erg s$^{-1}$} in 2016 and $1.2\times10^{34}$\,\mbox{erg s$^{-1}$} in 2018 (0.5--10\,keV, for the double-blackbody model and for 3.3\,kpc), which is still almost about ten times brighter than the quiescent level. The pulse profile displayed dramatic changes, apparently evolving from the complex multi-peak morphology observed in high-luminosity states to the more sinusoidal form characteristic of latency. The inspection of the X-ray light curves revealed two flares with unusual properties in the 2016 observation: they are long ($\sim$1\,ks to be compared with 0.1--1\,s of typical magnetar bursts) and faint ($\approx$$10^{34}$\,\mbox{erg s$^{-1}$}, with respect to $10^{38}$\,\mbox{erg s$^{-1}$} or more in magnetars). Their spectra are comparatively soft and resemble the hotter thermal component of the persistent emission. If the flares and the latter component have a common origin, this may be a spot on the star surface that is heated by back-flowing currents that are induced by a magnetospheric twist. In this hypothesis, since the increase in luminosity of \mbox{1E\,1613} during the flare is only $\sim$20\%, an irregular variation  of the same order in the twist angle could account for it.   }


  \maketitle

\section{Introduction}

The X-ray source \mbox{1E\,161348--5055} (\mbox{1E\,1613}) lies at the center of the young (2 kyr) 
supernova remnant (SNR) RCW\,103. It was discovered with \emph{Einstein} 
and was suggested to be a radio-quiet isolated neutron star (NS) by 
\citet{tuohy80}. Over the years, the behavior of \mbox{1E\,1613} has set it apart
 from any other compact object source 
\citep{deluca08,deluca17}: the source displays a strong variability on 
scales of months or years (a large outburst with an increase in 
luminosity from $\sim$$3\times10^{33}$ to over $3\times10^{35}$\,\mbox{erg s$^{-1}$}
occurred in 1999), and features a modulation in flux with a period 
of 6.67\,h, together with dramatic changes in the pulse profile (which are 
correlated with the flux level). Considering also the young age 
of \mbox{1E\,1613} and the lack of an optical or IR counterpart, 
\citet{deluca06} discussed two main possibilities: \mbox{1E\,1613} could be 
either a very young low-mass X-ray binary (LMXB, the first observed
inside an SNR, see also \citealt{bhadkamkar09}), or an isolated 
magnetar that slowly revolves at an abnormal period of 6.67\,h, possibly due to a propeller interaction with a fallback disk
(see also \citealt{li07,popov15}). Even more exotic pictures have been proposed, 
such as an LMXB with a supermagnetic NS locked in synchronous rotation 
with the orbit \citep{pizzolato08}, or an evolved Thorne--\.Zytkow 
object \citep{liu15}.

Recently, a remarkable event added new elements to the decade-long
enigma. On 2016 June 22, the Burst Alert Telescope (BAT) on board
the \emph{Neil Gehrels Swift Observatory} detected a short X-ray 
burst from the direction of RCW\,103
\citep{rea16,dai16}. The light curve of the short burst ($\sim$10\,ms) 
shows a double-peak profile, the spectrum of which is well described by a
blackbody model ($kT\sim 9$\,keV) and a luminosity of 
$\sim$$2\times10^{39}$\,\mbox{erg s$^{-1}$} (for a distance of 3.3\,kpc; \citealt{caswell75}; \citealt{rea16}). 
All in all, the event was a signature magnetar burst 
\citep{turolla15,kaspi17,esposito19}. Concurrently, an enhancement  by a factor 
$>$100 in the X-ray flux of \mbox{1E\,1613} with respect to the quiescent level (observed for several years and up to one month before)
was measured with the \emph{Swift} X-Ray Telescope (XRT). 
On \mbox{2016 June 25}, \emph{Chandra} and \emph{NuSTAR} 
observations showed a flux modulation  at the known 6.67\,h period with 
two main peaks per cycle, which is different from the nearly sinusoidal shape 
seen during the quiescent state of the source \citep{etdl11} and was detected 
for the first time \mbox{1E\,1613} at hard X-rays of up to $\sim$30\,keV 
\citep{rea16}. 

An intense multi-instrument monitoring showed that after 
more than 1\,yr since the onset of the outburst, \mbox{1E\,1613} was still $\text{about ten}$ 
times brighter than usual and that the total energy emitted 
($\sim$$2\times10^{42}$\,erg), the flux decay pattern, and the spectral
evolution were similar to what is generally observed in magnetars 
\citep{rea16,borghese18,cotizelati18}. Moreover, in  \emph{HST} images taken 
on 2016 July 4 and August 11, \citet{tendulkar17} detected a likely 
IR counterpart with $J$ magnitude 26.3 and 
$H$ magnitude 24.2, implying a minimum brightening of at 
least 1.3 magnitudes ($H$) compared to the non-detections in the 
previous (2002) observations \citep[with a different instrument,][]{dmz08} and an X-ray-to-IR luminosity ratio 
consistent with typical values or limits for magnetars and isolated NSs \citep{mignani11,olausen14}. 
Finally, the optical counterpart rules out for \mbox{1E\,1613} all binary 
scenarios in which the IR emission comes from an accretion disk or 
from a stellar companion (non-degenerate or white dwarf).

In this paper, we report on the results from the analysis of 
two \emph{XMM--Newton} exposures carried out in 2016 August and 2018 February that seized the source at two different flux levels along its decay pattern.
We also present VLT/NAOS+CONICA (NaCo) IR images taken at Paranal in the night of
2016 June 29--30, a few days before the \emph{HST} observations that discovered 
the counterpart of \mbox{1E\,1613} \citep{tendulkar17}.
We detected a faint source that can be identified to be that
of \citet{tendulkar17}, which was not detected in previous NaCo data that were collected in 2006 \citep{dmz08},
while \mbox{1E\,1613} was in quiescence.

\section{Very Large Telescope observations and results}

We performed a target of opportunity observation (Program ID: \mbox{297.D--5042(A)}) of the field of \mbox{1E\,1613} on
2016 June 30 at the ESO Paranal Observatory with NaCo
\citep{lenzen03,rousset03}, the adaptive optics imager and spectrometer
mounted at the VLT Unit\,1 (Antu). We adopted the
same setup as we used in our previous NaCo observations of \mbox{1E\,1613} (2006 May; Program ID: \mbox{077.D--0764(A)};
\citealt{dmz08}). The instrument was operated with the S27 camera, giving a
field of view of $28\arcsec \times28\arcsec$ and a pixel scale of
$0\farcs027$. The visual (VIS) dichroic element and wave-front sensor
($4500$--$10000$\,\AA) were used. Observations were performed in the $K_s$
filter ($\lambda=2.18$\,$\upmu \rm m$; $\Delta\lambda=0.35$\,$\upmu \rm m$
FWHM). The only suitable reference star for the adaptive optics correction
is the \mbox{GSC-2} star S230213317483 ($V\sim15.2$), located $21\farcs1$ away from
our target. We performed three observations, lasting 2,500\,s each, split into
sequences of short dithered exposures with 50\,s integration. At variance
with our 2006 observations, the target region was located close to the
center of quadrant 4 to keep it away from the malfunctioning quadrant
2\footnote{See the NaCo News Page at\\
\texttt{http://www.eso.org/sci/facilities/paranal/instruments/\ naco/news.html}.}
in the dithering pattern; this resulted in a $\sim$10$\arcsec$ shift of the
target within the field of view with respect to the 2006 data. Airmass was
in the 1.12--1.17 range; seeing conditions were very good, in the
$0\farcs30$--$0\farcs50$ range, mostly below $0\farcs40$. Sky conditions
were mostly photometric. Night (twilight flat fields) and day-time calibration frames (darks, lamp flat fields) were taken daily as part
of the NaCo calibration plan. The data were processed using the ESO NaCo
pipeline, and the science images were coadded using the \texttt{eclipse}
software \citep{devillard97} to produce a master image with 7,500 s exposure time.

Our goal is to search for sources in the error region of \mbox{1E\,1613} that display variability with respect to our 2006 observations. Thus, the source catalog produced by \citet{dmz08} was adopted as a reference for both astrometry and photometry. We ran a source detection  on our master image using the \texttt{SExtractor} software \citep{bertin96}. Using a set of $\sim$300 sources down to $K_s=20$, we superimposed the master image on the image obtained in 2006 with a root mean square (r.m.s.) $<0\farcs01$. Using the same set of stars, we fit the zero-point needed to convert the instrumental aperture magnitudes to $K_s$ magnitudes as computed in our previous work \citep{dmz08}. This exercise yields a rather large scatter in the flux measurements between the two epochs, with an r.m.s. of $\sim$0.25 mag. We noted that residuals clearly depend on the position of the sources within the field of view. This could be due to anisoplanatic effects (i.e., related to the change in point spread function as a function of distance from the guide star in adaptive optics observations) or to the positioning of the target in different quadrants. As a simple approach, we added a correction to the zero-point that was linearly dependent on the distance to the guide star. The fit results were much better, with an r.m.s. below 0.10 mag on the whole field of view. 

\begin{figure*}
\resizebox{\hsize}{!}{\includegraphics[angle=0]{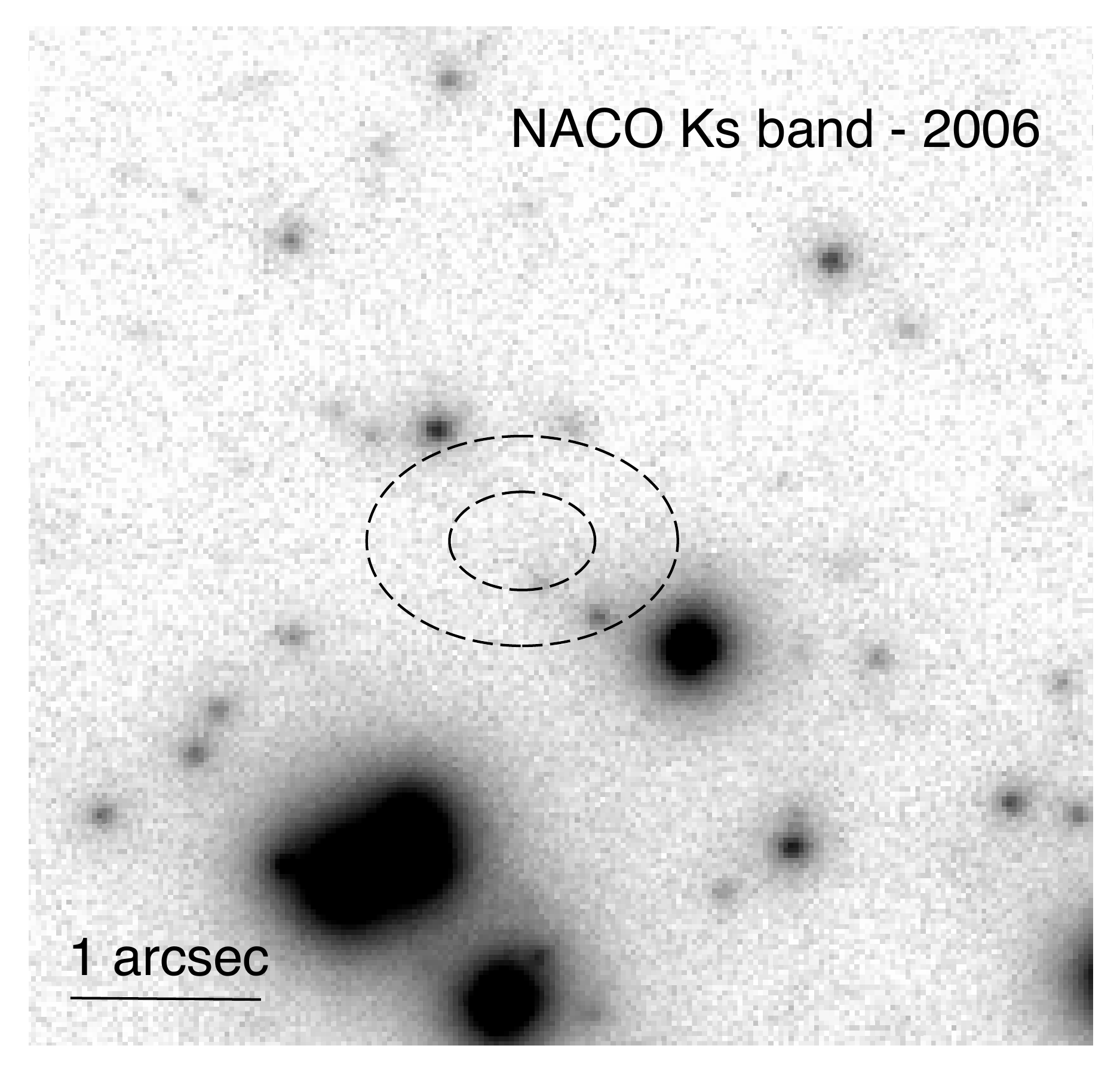}\includegraphics[angle=0]{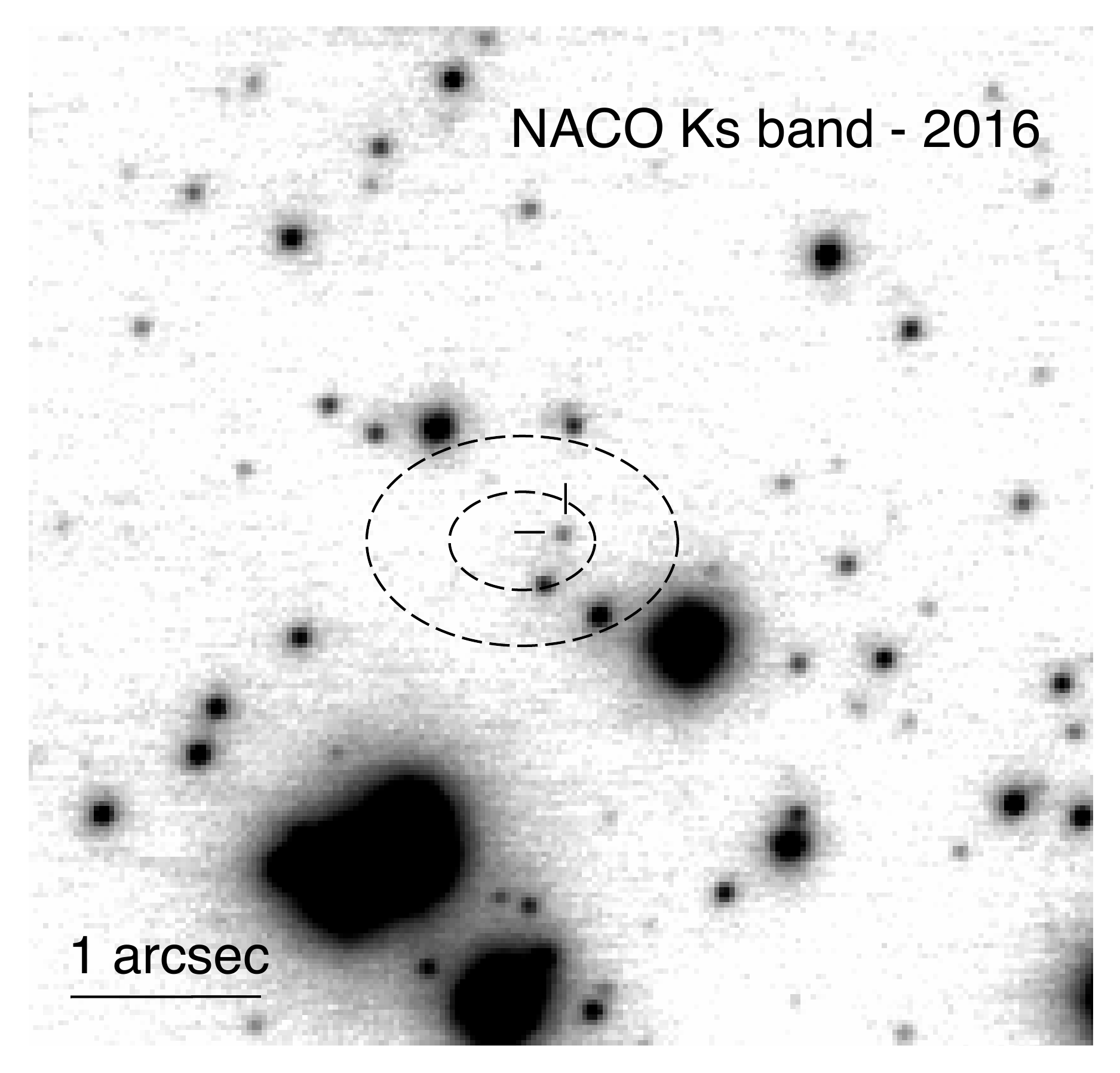}}
\caption{VLT/NaCo images of the field of \mbox{1E\,1613} in RCW\,103 in 2006 and
2016. The ellipses show the 68\% and 99\% position uncertainty of \mbox{1E\,1613}. The source detected
in 2016 at $K_s=20.68\pm0.12$ that previously was not detected ($K_s>21.2$) is marked with two ticks. 
The same source is labeled `8' in fig.\,2 of \citet{tendulkar17}.}
\label{counterpart}
\end{figure*}

Inspection of the master image shows a new source that is located close to
the center of the error region for \mbox{1E\,1613}. This was not detected in our 2006 data 
(see Fig.\,\ref{counterpart}). Its position is 
$\rm RA=16^{\rm h}17^{\rm m}36\fs21$, $\rm Decl.=-51^{\circ}02\arcmin24\farcs6$
(J2000), with an uncertainty of $\sim$$0\farcs1$ per coordinate 
(dominated by the uncertainty in the astrometric solution of the 2006
NaCo image that was used as a reference; \citealt{dmz08}) and its magnitude 
is $K_s=20.68\pm0.12$. 
All of the seven sources mentioned by \cite{dmz08} as candidate 
counterparts are also clearly detected in the new image, with no large 
flux variations with respect to the values measured in 2006 (r.m.s. of 
$\sim$0.13 mag). 
We performed simulations for which we added artificial sources to the 2006 NaCo 
image at the position of the new source detected in 2016. We estimate 
that a source at the 2016 magnitude level would have been easily 
detected, with a signal-to-noise ratio of $\sim$4.5. The $3\sigma$ 
upper limit to any undetected source at that position in the 2006 
image is $K_s>21.2$.

We also investigated a possible short-term variability of the new 
candidate counterpart of \mbox{1E\,1613}. To this aim, we generated images with 
1000\,s exposure time as well as with 2500\,s exposure time by 
combining 20 and 50 consecutive frames with 50\,s integration time, 
respectively. For each image, we performed astrometric and 
photometric calibration as described above for the master image, with 
similar accuracy. This allowed us to produce light curves with 
$\sim$1,000\,s and $\sim$2,500\,s binning for $\sim$380 and $\sim$900 
sources, respectively, including the new candidate counterpart as 
well as all of the seven candidate counterparts mentioned by 
\cite{dmz08}. As in our previous analysis \citep{dmz08}, we found a 
larger r.m.s. variability for fainter sources, which implies that our 
photometric measurements are contaminated by random errors  \cite[see 
discussion in][]{dmz08}. 
The new candidate counterpart ($K_s\sim20.7$) 
displays an r.m.s. of $\sim$0.22\,mag and of $\sim$0.23\, mag
on the 1,000\,s and on the 2,500\,s timescales, respectively. These 
results are broadly consistent with the apparent r.m.s. variability 
for sources in the $K_s$ magnitude range 20.0--21.0, which is of 
$\sim$0.18\,mag and of $\sim$0.13\,mag on the 1,000\,s (55 sources) and 
on the 2,500\,s timescale (335 sources), respectively. Thus, we 
cannot draw firm conclusions about a possible modulation of the flux 
of the new candidate counterpart. The same is true for all of the 
seven candidate counterparts mentioned by \cite{dmz08}.

\section{\emph{XMM--Newton} observations and results}\label{xraysection}
Two \emph{XMM--Newton} observations of \mbox{1E\,1613} were taken after the 2016 outburst, 
one on 2016 August 19--20 (Obs.ID~0743750201) and one on 2018 
February 14 (Obs.ID~0805050101). The 2016 exposure lasted about 
82.5\,ks, and the EPIC pn \citep{struder01} was operated in Small 
Window mode (time resolution: 5.7\,ms) and the MOS detectors
\citep{turner01} in Large Window mode (time resolution: 0.9\,s); all
cameras mounted the thin optical-blocking filter. The exposures of
the MOS cameras were interrupted after $\sim$0.5\,ks and later resumed,
with a gap of $\sim$3.2\,ks both in the MOS1 and  MOS2 data. The 2018 pointing was 63 ks long and all the 
detectors were in Full Frame mode (time resolution: 73.4\,ms for 
the pn and 2.7\,s for the MOSs), using the thin filter for the pn and the medium 
filter for the MOSs. The second observation was affected by intervals
of flaring particle background, which were removed using intensity 
filters on the light curves;
this screening reduced by $\sim$20\% the net exposure time in the 
pn and by a lesser amount in the MOS cameras (see Table\,\ref{data} for more details). 
\begin{table}
\caption{\emph{XMM--Newton} observations. In the second observation, the net exposure times are after the screening for proton flares.\label{data}}             
\centering                          
\begin{tabular}{llc}        
\hline\hline    
Obs.ID & Date & Net exposure (ks) \\   
 & & pn / MOS1 / MOS2 \\
\hline   
0743750201~(A) & 2016-08-19/20 & 56.7 / 76.9 / 76.7 \\      
0805050101~(B) & 2018-02-14 & 43.8 / 59.1 / 59.6 \\  
\hline                                   
\end{tabular}
\end{table}

The raw observation data files were processed with the Science Analysis
Software (\texttt{SAS};  \citealt{gabriel04}) v.17.1. To extract the 
event lists and spectra, we used the same regions as were selected in 
\citet{deluca06}: a circle with a radius of 15 arcsec centered on \mbox{1E\,1613}  and a circle with a radius of 
20 arcsec with a center at \mbox{$\rm RA=16^h17^m42\fs4$}, 
\mbox{$\rm Decl. = -51\degr02'38'' (J2000)$}, a region where the surface 
brightness of the RCW\,103 SNR is comparable to that of the surroundings of the source, to 
estimate the background. Spectra were rebinned so as to have a minimum 
of 30 counts per energy bin, and for each spectrum, the response matrix 
and ancillary files were generated with the \texttt{SAS} tasks 
\texttt{rmfgen} and \texttt{arfgen}. To convert the event times into the  
solar system barycenter, we used the task \texttt{barycen}.

The evolution of the flux, the spectral shape, and the pulse 
profile of \mbox{1E\,1613} in the aftermath of the outburst have been 
studied in detail by \citet{borghese18} and \citet{cotizelati18} 
over a period encompassing the two \emph{XMM--Newton} observations. Here, we
concentrate on the individual data sets.
\begin{figure}
\centering
\resizebox{\hsize}{!}{\includegraphics[angle=0]{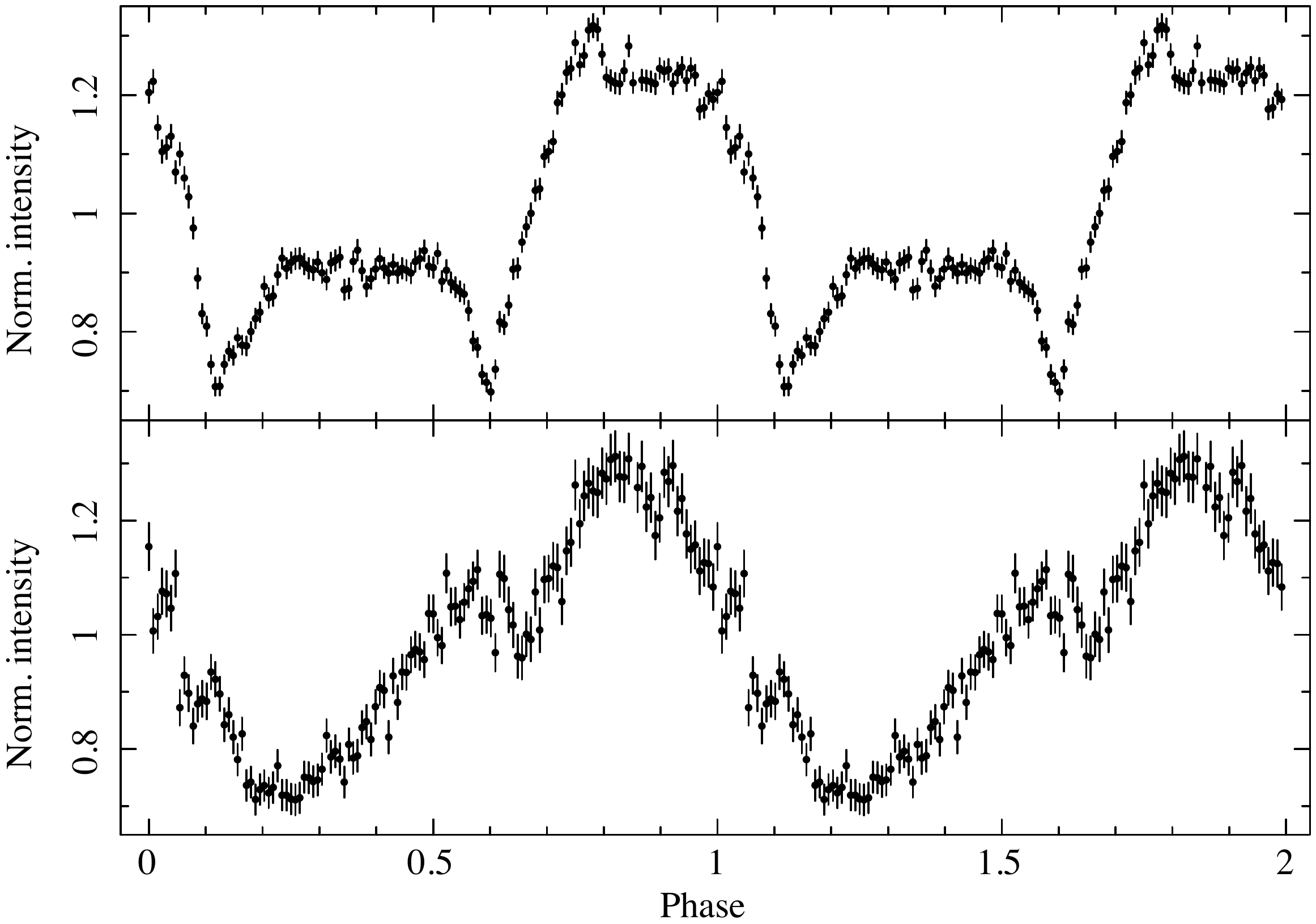}}
\caption{Folded light curves of \mbox{1E\,1613}. The top panel shows the 2016 data set (A) and the bottom panel the 2018 data set (B). The energy range is 1--10\,keV, and both data sets were folded with epoch MJD\,57619 in the \texttt{FTOOLS} task \texttt{efold} and the period of \citet[][$P=24\,030.42$\,s]{etdl11}. The profile apparently evolves from the complex multi-peak structure observed in high flux states to the simpler (nearly sinusoidal) shape observed in quiescence (see fig.\,2 of \citealt{deluca06}, fig.\,3 of \citealt{rea16}, and fig.\,4 of \citealt{etdl11} for a collection of the pulse profiles that were observed in quiescence).
 \label{efold}}
\end{figure}

The 6.67 h flux modulation is evident in both observations and apparently evolves from the complicate shape that is characteristic of the 
outbursts toward the nearly sinusoidal profile observed in quiescence 
(Fig.\,\ref{efold}; see also fig.\,2 of \citealt{deluca06} and fig.\,4 of \citealt{etdl11}). In the first observation, the high flux combined 
with the long exposure resulted in spectra with hundreds of thousands of 
source counts ($>$$2\times10^5$ in the pn and $>$$10^5$ in each MOS). 
In spectra of such high statistical quality, systematic (calibration) 
errors are important in the error budget, as can be seen by inspecting
the fit residuals at the energies of the 
effective area edges, where large variations occur rapidly. For this 
reason we added to the spectra an energy-independent systematic 
uncertainty of 3\% \citep[see, e.g.,][]{molendi09}.

\begin{figure*}
\centering
\resizebox{\hsize}{!}{\includegraphics[angle=0]{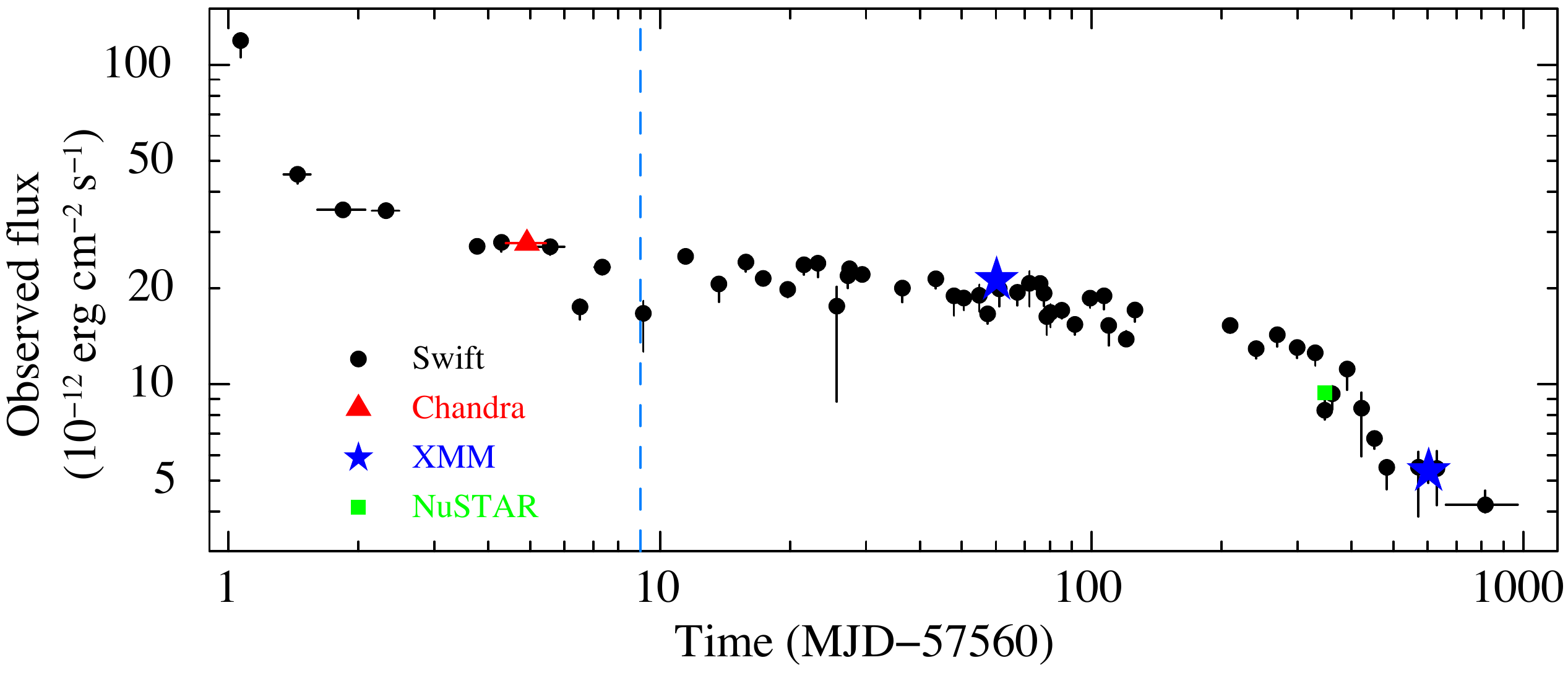}}
\caption{Long-term light curve of \mbox{1E\,1613} starting from the 2016 burst. 
The flux (not corrected for the absorption) is in the 0.5--10\,keV band. 
Except for the \emph{XMM--Newton} points, the data are from the Magnetar Outburst 
Online Catalog (MOOC, \texttt{http://magnetars.ice.csic.es/}; 
\citealt{cotizelati18}) and the flux values were obtained from a 
double-blackbody fit. For some data points, the error bar 
is smaller than the symbol. The dashed line indicates the epoch of 
the VLT observation.\label{lightcurve}}
\end{figure*}

The spectra may be modeled by a double blackbody or a 
blackbody-plus-power-law model and an absorption component. For 
the latter, we adopted the abundances by \citet{wilms00} and the 
T\"ubingen-Boulder interstellar medium (ISM) absorption model. The parameters derived 
from the spectral fits are similar to those from previous analyses
\citep{deluca06,rea16,borghese18} and are 
summarized in Table\,\ref{fits}.
\begin{table*}
\begin{tiny}
\caption{Summary of the spectral results.} \label{fits}
\centering
\begin{tabular}[c] {lccccccccc}
\hline\hline
Obs. & Model & $N_{\rm H}$ & $k_B T_1$ & $R_{\mathrm{bb1}}$\tablefootmark{a} & $\Gamma$/$k_B T_2$ & $R_{\mathrm{bb2}}$\tablefootmark{a} & Flux\tablefootmark{b} & Luminosity\tablefootmark{b} & $\chi^{2}_{\nu}$\,(dof) \\
& &  ($10^{22}$ $\rm cm^{-2}$) & (keV) & (km) & (--/keV) & (km) & ($10^{-11}$\,\mbox{erg cm$^{-2}$ s$^{-1}$}) & ($10^{34}$\,\mbox{erg s$^{-1}$}) & \\
\hline
A & BB+BB  & $1.10\pm0.01$ & $0.554^{+0.003}_{-0.004}$ & $1.89^{+0.02}_{-0.03}$ & $1.20\pm0.03$ & $0.15\pm0.01$ & $2.118\pm0.006$ & $4.80\pm0.03$ & 1.07\,(1803)\\
 & BB+PL & $1.32^{+0.03}_{-0.04}$ & $0.574\pm0.002$ & $1.68\pm0.01$ & $2.80^{+0.06}_{-0.07}$ & -- & $2.126^{+0.006}_{-0.007}$ & $6.56^{+0.06}_{-0.05}$ & 1.15\,(1803)\\
 B & BB+BB & $0.98\pm0.02$ & $0.527^{+0.006}_{-0.007}$ & $1.04^{+0.03}_{-0.02}$ & $1.14\pm0.06$ & $0.09^{+0.01}_{-0.02}$ & $0.529^{+0.003}_{-0.002}$ & $1.20\pm0.01$ & 1.07\,(1255)\\
 & BB+PL & $1.32^{+0.08}_{-0.07}$ & $0.543\pm0.003$ & $0.92^{+0.01}_{-0.02}$ & $3.1^{+0.1}_{-0.2}$ & -- & $0.532^{+0.003}_{-0.002}$ & $2.01^{+0.04}_{-0.03}$ & 1.09\,(1255)\\
 \hline
\end{tabular}
\tablefoot{Uncertainties are given at the 1$\sigma$ confidence level.\\
\tablefoottext{a}{Radius at infinity assuming a distance of 3.3\,kpc.}\\
\tablefoottext{b}{Observed (not corrected for the absorption) flux in the 0.5--10\,keV range. The luminosity is in the 0.5--10\,keV range  for a distance of 3.3\,kpc.}}
\end{tiny}
\end{table*}
Figure\,\ref{lightcurve} shows the long-term evolution of the observed flux (the \emph{XMM--Newton} data refer to the double-blackbody model).

\begin{figure}
\centering
\resizebox{\hsize}{!}{\includegraphics[angle=0]{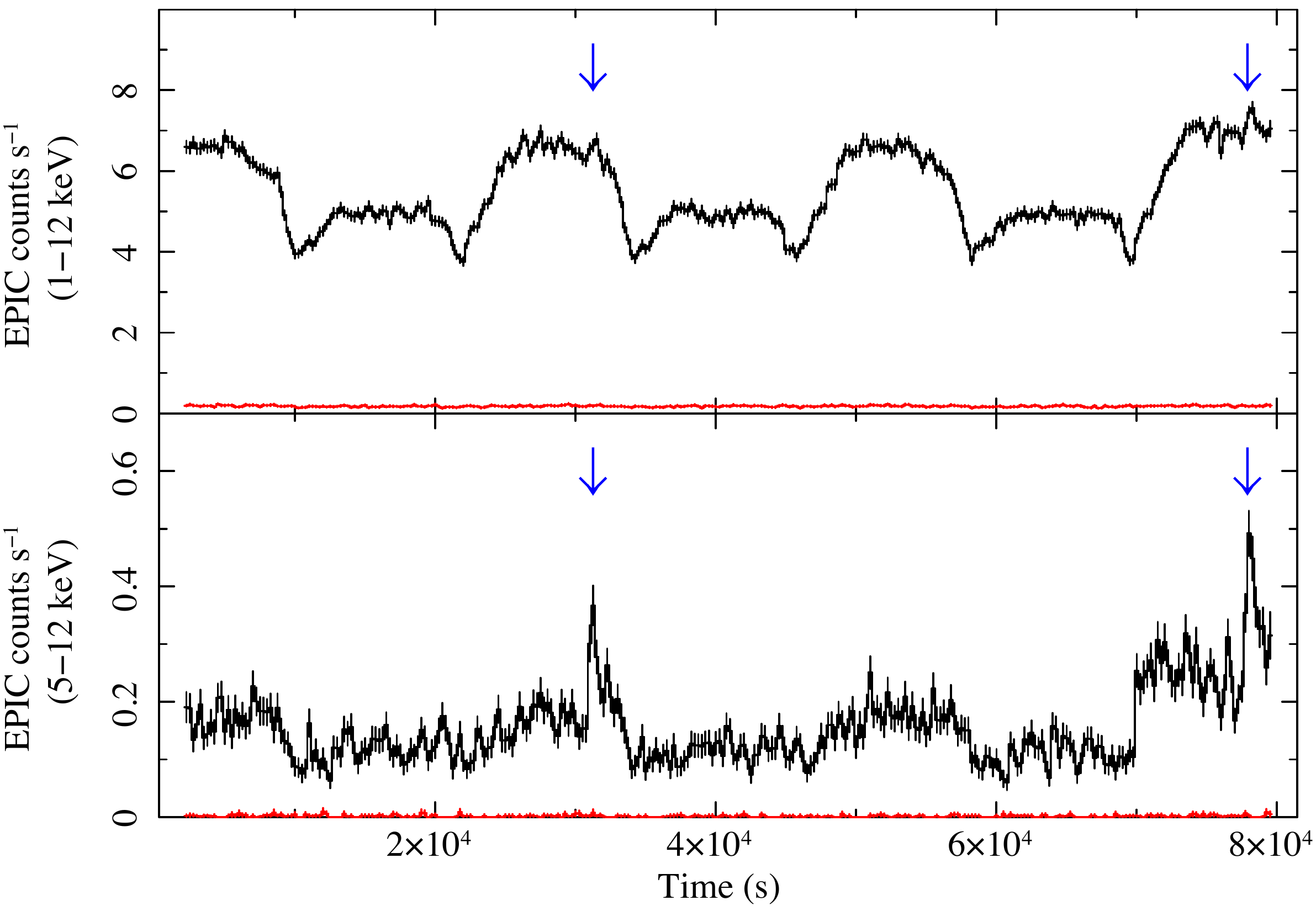}}
\caption{Light curves of \mbox{1E\,1613} from the 2016 observation (A) in the 1--12\,keV range (top)
and in a hard band (5--12\,keV). The time bin is 250\,s. The background contribution is negligible and 
was not subtracted, but the background level in the proper energy band is shown in red in each panel. The arrows indicate the faint flares (more 
prominent in the hard band) that are discussed in Section\,\ref{xraysection}. 
\label{flares}}
\end{figure}

Inspecting the light curves, we realized that they contained unusual features
in the first observation. At least two long-lasting flares occurred, the first
$\sim$32\,ks after the start of the observation, the second at about 78\,ks
(see Fig.\,\ref{flares}). They can be identified because ($>$5\,keV) they appear like spikes in the hard band. The first flare lasted approximately 
1.2\,ks and had a faster rise than decline; the second flare had a slightly more symmetric
profile and lasted $\sim$2.2\,ks. No similar events were found in the second observation.

\begin{figure}
\centering
\resizebox{\hsize}{!}{\includegraphics[angle=0]{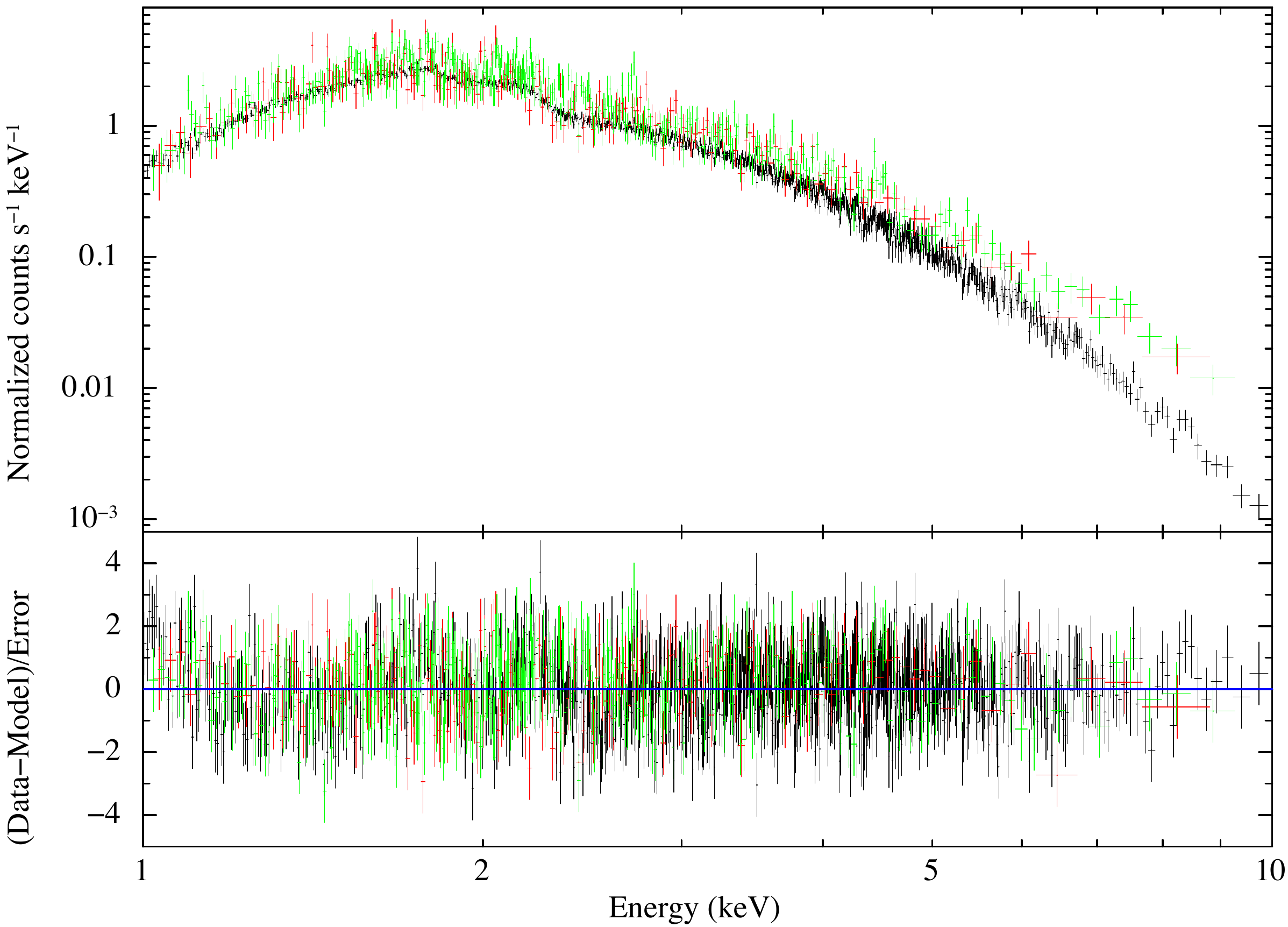}}
\caption{Spectra of the flares (the data of the first are plotted 
in red, those of the second in green) compared with that of the 
average emission (in black). The model fit to the data is the 
double blackbody. For clarity, we plot only the 
pn points. Bottom: Post-fit residuals. \label{spec_flares}}
\end{figure}

Extracting the spectra from the time intervals of the flares, we found that the 
spectral shapes of the two flares are very similar. As a first guess, we assumed that the
flare emission is superimposed on the persistent emission, 
and we fit the flare spectra with a model consisting of the average 
spectrum of observation A plus an additional component, either a power law 
or a blackbody.
The fits are essentially equally good; the additional component 
resembles the hard component of the average spectrum. In the 
blackbody fit, the temperature  converges to a 
value that is compatible with the value of the hot blackbody in Table\,\ref{fits}.
In other words, the spectra of the flares are extremely similar to the spectrum 
of the persistent emission with an stronger hot component. 
This can be visually assessed in  Fig.\,\ref{spec_flares}, where we show
a simultaneous fit of the spectra of the average emission and of the 
two flares where all parameters are tied except for the hot 
blackbody normalization. The resulting radii are $0.25^{+0.01}_{-0.02}$ 
and $0.27^{+0.02}_{-0.01}$\,km for the first and second flare, 
respectively ($\chi^2_\nu=1.10$ for 2756 degrees of freedom, dof). 
The corresponding luminosities are $(5.7\pm0.1)\times10^{34}$ and
$(6.1\pm0.1)\times10^{34}$\,\mbox{erg s$^{-1}$} (0.5--10\,keV), that is, only 
$\approx$20--30\% more than the average luminosity (but we note that both 
events occur at the maximum of the pulse profile, although not exactly at the same phase: they are separated by $\sim$1.9 cycles).  
\section{Discussion}
\begin{figure}
\centering
\resizebox{\hsize}{!}{\includegraphics[angle=0]{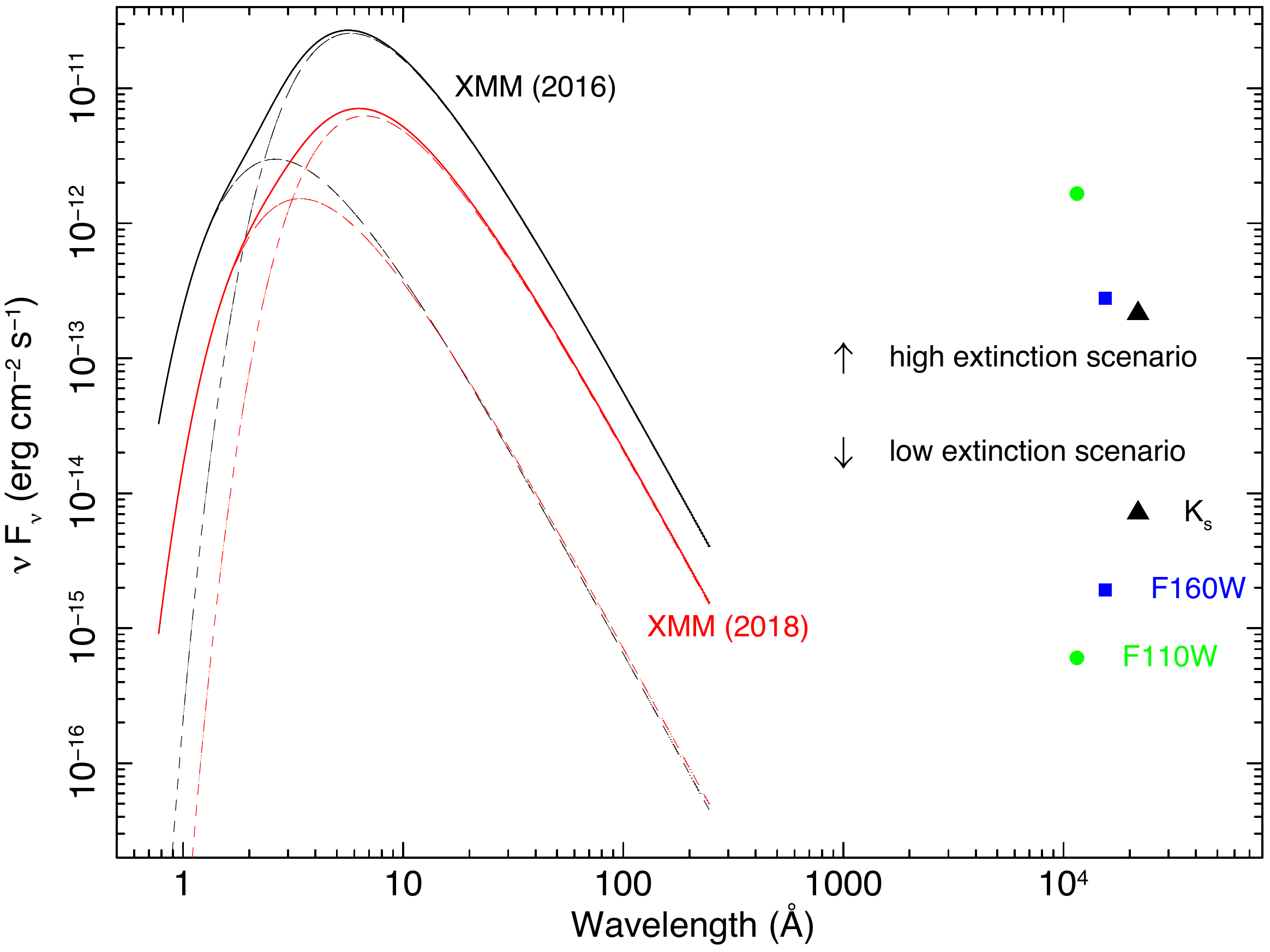}}
\caption{Spectral energy distribution for \mbox{1E\,1613} from X-rays to 
near IR. The \emph{XMM--Newton} spectral
models (double blackbody) are plotted in black for the 2016 data and in 
red for the 2018 data; the two blackbody components are plotted with the 
same colors and dashed lines. The black triangles indicate the June 2016
VLT values in the case of low and high extinction ($A_V$ from 3.6 to 36). The
\emph{HST} points are from \citet{tendulkar17} and the exposures were taken
in August 2016 (similar values were found earlier, in July 2016).\label{sed}}
\end{figure}

In 2016, after hibernating for more than a decade, \mbox{1E\,1613} entered a new phase 
of magnetar-like activity, which produced the first outburst from this source. The source was monitored intensely and was also followed in hard X-rays and at
IR wavelengths \citep{rea16,dai16,tendulkar17,borghese18}. 
All studies spotted strong similarities between this unusual source and
magnetars, which underlines the possibility that the compact
object might be a magnetar \citep{deluca06}. The space for a binary system was already narrow
in view of the limit on the IR emission that was previously obtained by 
\citet{dmz08}, but the ignition of \mbox{1E\,1613} in the IR that was observed with \emph{HST} by
\citet{tendulkar17} strongly argues against the possibility of accretion
onto the neutron star. The \emph{HST} detection implies a brightening 
$>$1.3\,mag in the F160W filter band ($H$ band, $\lambda=1.545$\,$\upmu \rm m$;
$\Delta\lambda=0.29$\,$\upmu \rm m$ FWHM) and $>$1.8\,mag in the F110W filter band (wide $J$, $\lambda=1.15$\,$\upmu \rm m$;
$\Delta\lambda=0.5$\,$\upmu \rm m$ FWHM). Our (non-simultaneous) VLT
observation implies an increase in flux density of at least  $\sim$0.5\,mag in the $K_s$ band. 
The extinction toward \mbox{1E\,1613} is uncertain. The dust reddening mapped 
in its direction is $A_V=36$ \citep{schlafly11}, and high extinction 
is indicated also by the colors of field stars \citep{dmz08}. On 
the other hand, much lower values are obtained from optical and IR
observations of the SNR and from the {$N_{\rm H}$}\ measured in the X-ray 
observations \citep{oliva89,tendulkar17}. Following \citet{tendulkar17}, 
we assume an $A_V$ from 3.6 to 36 but, as they note, a low $A_V$
 seems more likely because it is suggested by measurements of \mbox{1E\,1613} and 
its SNR, while the high values come from analyses of a vast field and unrelated stars. The IR-to-X-ray spectral energy distribution of \mbox{1E\,1613} for 
the extreme values of $A_V$ is shown in Fig.\,\ref{sed} (which, we stress,
must be taken with caution because the source is variable and the 
exposures are not simultaneous).  
Propeller interaction with a surrounding fallback disk is often proposed
as a possible explanation of the slow rotation of \mbox{1E\,1613} \citep{deluca06,li07,tong16,ho17}. Given the lack of constraints on the presence of pulsations in the VLT data, it is hard to tell whether the IR emission is of magnetospheric origin or is due to a disk heated by the enhanced emission of \mbox{1E\,1613} during 
the outburst. In the case of low extinction, the $K_s$-to-X-ray flux ratio 
is $\sim$$1.8\times10^{-5}$, which is similar to the corresponding \emph{HST} 
ratios and is in the range observed for magnetars and isolated neutron 
stars \citep{mignani11,olausen14}. With this assumption, the detection in itself therefore does 
not support the presence of a fallback disk around the star.

A careful analysis of the source light curves in X-rays revealed the 
presence of two flares with properties much at variance with those 
of canonical magnetar bursts. They last much longer ($\sim$1\,ks vs. 
$0.1$--$1$\,s), are less energetic (\mbox{$L\approx 10^{34}$\,\mbox{erg s$^{-1}$}}, exceeding 
the persistent luminosity by only $\sim$20\%, vs. \mbox{$L\ga10^{38}$\,\mbox{erg s$^{-1}$}}), 
and are softer, with a spectrum quite close to the hotter thermal 
component of the persistent flux. Because of their long duration, however, their fluence is rather high, $\approx$$7\times10^{-9}$ and $2\times10^{-8}$\,\mbox{erg cm$^{-2}$} for the first and second event, respectively, and comparable with standard short burst from magnetars \citep[e.g.,][]{aptekar01,gmm06,gogus00}.

To the best of our knowledge, an 
event like this has not been observed before in a magnetar or in 
other classes of X-ray emitting isolated NSs. However, flares similar 
to those reported here might well have occurred in other sources and passed
unnoticed. The strong similarity of the flare spectrum with the hotter 
thermal component seen in the persistent emission suggests a common origin.
In magnetars, the latter is believed to come from a limited region on the 
star surface that is heated by dissipative processes. If the flares observed in 
\mbox{1E\,1613} do come from the same hot spot, they would be hardly visible in a 
source with a period ($\approx$0.1--10\,s) much shorter than the 
flare duration ($\approx$1\,ks), as is the case of all the 
known X-ray pulsars. In this sense, \mbox{1E\,1613} is unique because of its 
period, which is much longer than the flare duration. Further pursuing the idea 
that the flares originate from a spot on the star surface, they could 
be produced by some irregularities in the process responsible for the 
heating. Although the nature of such processes in magnetars is still 
debated \citep[see, e.g.,][for recent reviews]{turolla15,kaspi17,gourgouliatos18}, we 
mention that if the star surface is heated by backflowing currents, 
an increase in luminosity by $\sim$20\% can easily be explained 
by a change of the same order in the twist angle or in the potential 
drop that accelerates the charges \citep{beloborodov09}. 
We also note that the second flare occurred approximately at the same rotational phase as the first, after $\sim$1.9 cycles. Although with just two events a coincidence cannot be excluded, this might indicate that they came from the same hot spot.

\begin{acknowledgements}
The scientific results reported in this article are based on 
observations obtained with \emph{XMM--Newton}, an ESA science mission with 
instruments and contributions directly funded by ESA Member States 
and NASA. Based on observations collected at the European Organisation for Astronomical Research in the Southern Hemisphere under ESO programmes \mbox{077.D--0764(A)} and \mbox{297.D--5042(A)}. PE and ADL acknowledge funding in the framework of the 
project ULTraS (ASI--INAF contract N.\,2017-14-H.0). FCZ is supported by a Juan de la Cierva fellowship.
\end{acknowledgements}

\bibliographystyle{aa} 
\bibliography{biblio} 

\end{document}